\pdfoutput=1
\documentclass[11pt,bezier,amstex]{article}

\usepackage[in]{fullpage}
\usepackage{times}  
\usepackage{helvet}  
\usepackage{courier}  
\usepackage[hyphens]{url}  
\usepackage[square,sort,comma,numbers]{natbib}
\usepackage{caption} 
\DeclareCaptionStyle{ruled}{labelfont=normalfont,labelsep=colon,strut=off} 
\usepackage{algorithm}
\usepackage{algorithmic}
\usepackage{bmpsize}
\usepackage[pdftex]{graphicx}
\usepackage{amsmath,amssymb,amsfonts}
\usepackage{makecell}
\usepackage{multirow}
\usepackage{multicol}
\usepackage{hhline}
\usepackage{subfigure}
\usepackage{url}
\usepackage{lipsum} 
\usepackage[suppress]{color-edits}

\addauthor{xl}{magenta}
\addauthor{sh}{olive}
\addauthor{bc}{cyan}
\addauthor{ab}{blue}
\addauthor{td}{orange}
\addauthor{lt}{green}

\newcommand{\beq}{\begin{equation}}
\newcommand{\eeq}{\end{equation}}

\newcommand\R{\mathbb{R}}

\newcommand\0{\mathbbm{0}}

\newcommand{\x}{\mathbf{x}}
\newcommand{\y}{\mathbf{y}}

\newcommand{\vertiii}[1]{{\left\vert\kern-0.25ex\left\vert\kern-0.25ex\left\vert #1
    \right\vert\kern-0.25ex\right\vert\kern-0.25ex\right\vert}}

\newcommand{\norm}[2]{\ensuremath \|#1\|_{#2}}

\DeclareMathOperator{\argmin}{argmin}

\newcommand {\commentout}[1] {}

\def\ints{{{\rm Z} \kern -.35em {\rm Z} }}  
\def\smallints{{{\rm Z} \kern -.3em {\rm Z} }}  
\def\pints{{{\rm I} \kern -.15em {\rm N} }}      
\newcommand{\reals}{\mathbb R}

\def\cplx{{{\rm I} \kern -.45em {\rm C} }}       
\def\l2{\rm {\mathcal L}^{2}(\reals)}            

\renewcommand{\norm}[1]{\lVert#1\rVert}

\newcommand{\be}{\begin{eqnarray}}
\newcommand{\ee}{\end{eqnarray}}
\newcommand{\bea}{\begin{eqnarray}}
\newcommand{\eea}{\end{eqnarray}}
\newcommand{\beaa}{\begin{eqnarray*}}
\newcommand{\eeaa}{\end{eqnarray*}}
\newcommand{\bnad}{\begin{nad}}
\newcommand{\enad}{\end{nad}}

\usepackage{newfloat}
\usepackage{listings}
\floatstyle{ruled}
\newfloat{listing}{tb}{lst}{}
\floatname{listing}{Listing}
\usepackage{authblk}
\title{Learning and Dynamical Models for Sub-seasonal Climate Forecasting: Comparison and Collaboration}

\author[1,3]{Sijie~He}
\author[1]{Xinyan~Li}
\author[2]{Laurie~Trenary}
\author[2]{Benjamin~A~Cash}

\author[2]{Timothy~DelSole}
\author[3]{Arindam~Banerjee}
\affil[1]{Department of Computer Science \& Engineering,  University of Minnesota, Twin Cities}
\affil[2]{Department of Atmospheric, Oceanic, and Earth Science, George Mason University}
\affil[3]{Department of Computer Science, University of Illinois Urbana-Champaign}
\affil[ ]{hexxx893@umn.edu, lixx1166@umn.edu, ltrenary@gmu.edu, bcash@gmu.edu, tdelsole@gmu.edu, arindamb@illinois.edu}

\date{}
\usepackage{bibentry}

\begin{document}

\maketitle

\begin{abstract}
Sub-seasonal climate forecasting (SSF) is the prediction of key climate variables such as temperature and precipitation on the 2-week to 2-month time horizon. Skillful SSF would have substantial societal value in areas such as agricultural productivity, hydrology and water resource management, and emergency planning for extreme events such as droughts and wildfires. Despite its societal importance, SSF has stayed a challenging problem compared to both short-term weather forecasting and long-term seasonal forecasting. Recent studies have shown the potential of machine learning (ML) models to advance SSF. In this paper, for the first time, we perform a fine-grained comparison of a suite of modern ML models with start-of-the-art physics-based dynamical models from the Subseasonal Experiment (SubX) project for SSF in the western contiguous United States. Additionally, we explore mechanisms to enhance the ML models by using forecasts from dynamical models. Empirical results illustrate that, on average, ML models outperform dynamical models while the ML models tend to be conservatives in their forecasts compared to the SubX models. Further, we illustrate that ML models make forecasting errors under extreme weather conditions, e.g., cold waves due to the polar vortex, highlighting the need for separate models for extreme events. Finally, we show that suitably incorporating dynamical model forecasts as inputs to ML models can substantially improve the forecasting performance of the ML models. The SSF dataset constructed for the work, dynamical model predictions, and code \footnote{The SSF dataset is publicly available at \url{https://sites.google.com/view/ssf-dataset}. The codebase can be found at \url{https://github.com/Sijie-umn/SSF-MIP}.} for the ML models are released along with the paper for the benefit of the broader machine learning community.

\end{abstract}

\section{Introduction}
\label{sec: intro}



Nowadays, weather forecasts are routinely available out to a few days, and seasonal forecasts are routinely available out to a few months. These forecasts are based largely on dynamical models that solve partial differential equations (PDEs) derived from the laws of physics. On the other hand, sub-seasonal climate forecasting (SSF), which refers to the prediction of key climate variables, e.g., temperature and precipitation on 2-week to 2-month time scales, are not yet routined. Nevertheless, two high-profile reports from the National Academy of Sciences (NAS) discuss the immense societal values of SSF \cite{nati10, boar16}. In particular, skillful SSF in the western contiguous United States would allow for better hydrology and water resource management, agriculture, and emergency planning for extreme events such as droughts and wildfires \cite{white2017potential}. Currently, statistical forecasts of week 3-4 are available from the National Weather Service \cite{nws}, but the precipitation forecasts still are characterized as ``experimental'' rather than ``operational''. Sub-seasonal forecasts based on dynamical models are available every week through the Subseasonal Experiment (SubX) project, but the full utility of these forecasts for operational forecasting still remains to be determined \cite{subX}.

SSF is challenging for a variety of reasons. First, high-quality SSF has proven difficult to accomplish compared to both short-term weather forecasting and long-term seasonal forecasting \cite{vitart2012}. Due to the chaotic nature of atmosphere, weather events can not be accurately predicted beyond two weeks using dynamical models \cite{lorenz1963deterministic}. From a physical point of view, the predictability on sub-seasonal time scales depends on correctly modeling the atmosphere, ocean, and land, including their interactions and couplings as well as the memory effects of land and ocean. In addition to these physical complexities, SSF poses unconventional time series prediction problems. Given a training set $\{x_{1:t},y_{1:t} \}$, where $y$ denotes the target response variable, e.g., land temperature, and $x$ denotes suitable covariates, typical temporal models, ranging from auto-regressive models \cite{lutkepohl2005new} to recurrent neural networks \cite{gers1999learning, sutskever2014sequence}, focus on predicting $y_{t+1}$ or maybe $y_{t+1:t+\tau}$ for small $\tau$. Instead, SSF is about predicting $y_{T+1:T+\tau}$ for large $T \gg t$, e.g., weather prediction one month down the line. The long temporal range along with the nonlinear dynamics and complex interactions makes SSF challenging.

For climate forecasting, one standard baseline for comparing forecasts is the so-called climatology \cite{trewin2007role}, i.e. the 30-year average temperature/precipitation for each calendar day at each geographic location. Despite its simplicity, climatology provides a competitive benchmark for SSF. For instance, in the last Forecast Rodeo \cite{rodeo}, a SSF competition sponsored by the U.S. Bureau of Reclamation and the NOAA/National Integrated Drought Information System \cite{rodeo_2}, about half of the submitted forecasts could not beat climatology. Thus, for any more advanced SSF models, the first order of business is to do better than climatology. Recently, progress has been made in developing ML models \cite{hwang2019improving, heli21, weyn2021sub, srju21} which have shown great promise for outperforming climatology.

In this paper, we consider two new directions for SSF: first, comparing and contrasting ML models for SSF with an arguably stronger baseline provided by physics-based dynamical models; 
and second, exploring enhancing the ML models by using forecasts from such dynamical models. For the comparison, earlier literature has done such comparisons with certain statistical approaches and has illustrated dynamical models to have better forecasting ability \cite{barnston2012skill}. Instead, we do the comparison with a suite of modern ML methods, including non-parametric AutoKNN \cite{hwang2019improving}, multitask Lasso \cite{lasso,jalali2013dirty}, gradient boosted trees \cite{friedman2001greedy,xgboost}, and deep encoder-decoder networks \cite{heli21}, and illustrate that on average ML models outperform dynamical models on SSF. With considerably more details, our empirical analysis demonstrates key properties of ML-based vs. dynamical model-based predictions. In particular, most ML models generate conservative forecasts with small values, whereas dynamical models are more aggressive, generating forecasts with large scale. So when dynamical models are wrong, they can be wrong by a large amount; on the flip side, when dynamical models are correct, they can be more accurate than ML models. Further, we illustrate that ML models make most of their bad predictions during extreme events, e.g., unusual cold waves in North America, for which there is not enough training data. More practically, these results suggest that a separate ML model for extreme events will potentially help improve aggregate performance. The second direction is using physics-based dynamical model forecasts as covariates in the ML models. We show that using dynamical model forecasts as inputs improves the ML model forecasts, and the improvements are statistically significant.

We briefly emphasize the contributions of our work. We are {\em not} proposing another new algorithm which improves performance on an existing task using benchmark datasets, such as MNIST \cite{mnist} and ImageNet \cite{imagenet}. We are enabling a new application area for machine learning based on one of the most challenging and societally important scientific problems in the context of climate forecasting. We are reporting promising results using ML models, reporting detailed and nuanced empirical analysis acknowledging the strengths of both ML and dynamical models, and illustrating gains by using dynamical model forecasts as covariates in the ML models. We also suggest ways of further improving the ML models, e.g., by separately modeling extreme events. Finally, the dataset constructed for this work, dynamical model predictions, and code for the ML models are released along with the paper to replicate and hopefully extend our work. We also hope that the SSF dataset will become a standard benchmark like MNIST \cite{mnist} or ImageNet \cite{imagenet}, and accelerate advances on SSF.

\section{Related Work}
\label{sec: related_work}



\textbf{Dynamical models and S2S forecasting.} Nowadays, weather predictions rely heavily on ensemble forecasts from physics-based dynamical models~\cite{barnston2012skill}. On sub-seasonal to seasonal (S2S) time scales, forecasts have shown limited predictive skill compared to the climatology \cite{vitart2004monthly,vitart2014evolution,weigel2008probabilistic}. However, successful S2S predictions can be performed for certain regions and seasons \cite{li2015evaluation,delsole2017predictability}, as well as certain climate states \cite{mariotti2020windows}. In order to better understand the conditions that lead to enhanced predictability and to improve sub-seasonal forecasts, projects such as S2S \cite{vitart2017subseasonal} and SubX \cite{subX} have been established. These coordinated multi-model efforts act both to fulfill the growing needs of real world applications and to enrich our understanding of S2S prediction and predictability.

\textbf{ML on weather and S2S forecasting.} Recently, increasing efforts have been made to tackle complex problems in climate science using machine learning. Such applications aim to advance weather forecast skill using deep learning methods \cite{liu2016application,ham2019deep,dueben2018challenges,scher2019weather}. Despite early studies that show dynamical models outperform statistical models for ENSO seasonal forecasts \cite{barnston2012skill}, recent advances in machine learning, especially the development of deep learning, are making the performance of ML models more competitive with dynamical models for both weather \cite{grover2015deep,shi2017deep,dueben2018challenges} and seasonal \cite{stevens2021graph} prediction.


In particular, machine learning models have started to be used to improve forecast skills for predictions of temperature, precipitation, and other climate variables on sub-seasonal time scales \cite{hwang2019improving,heli21,weyn2021sub,srju21}. Some successful ML approaches for S2S forecasting include \cite{hwang2019improving} and \cite{heli21}, where both works show increased predictive skill for ML models compared to climatic baselines, e.g., climatology and damped persistence. Such advances from ML models are particularly relevant and valuable because dynamical models have limited predictive skills at sub-seasonal time scales \cite{predictability_s2s}.

\section{Sub-seasonal Climate Forecasting}
\label{sec: ssf}
\subsection{Problem Statement}
We focus on forecasting temperature anomalies over days 15 - 28, i.e., predicting average temperatures anomalies 2 weeks ahead of time, over the western contiguous U.S., which follows the Forecast Rodeo competition \cite{rodeo_2}. The spatial region is bounded by latitudes 25N-50N and longitudes 93W-125W at 1$^{\circ}$ latitude by 1$^{\circ}$ longitude spatial resolution with 508 grid points. The temporal range of interest and temporal resolutions are determined by each SubX model and its initialization frequency (see Table~\ref{table: data-subx}).

\subsection{Ground Truth Dataset}

The ground truth dataset is constructed from NOAA's Climate Prediction Center (CPC) Global Gridded Temperature dataset \cite{fan2008global}, which contains observations from the Global Telecommunication System (GTS) gridded using the Shepard Algorithm~\cite{Shepard}. Commonly applied for forecast verification by NOAA/CPC \cite{fan2008global}, the CPC dataset provides daily maximum and minimum 2m temperatures (air temperature at 2 meters above the surface - tmp2m) at 0.5 $^\circ$ by 0.5 $^\circ$ spatial resolution from Jan 1, 1979 to present.


To obtain the ground truth temperature anomalies for weeks 3 \& 4, we preprocess the data as follows: (1) daily 2m temperature at each grid point is taken as the average of daily maximum and minimum tmp2m, (2) all missing values are imputed by averaging the daily tmp2m of its spatial/temporal neighbors, (3) the tmp2m at 0.5 $^\circ \times 0.5^\circ$ resolution are linearly interpolated to a 1 $^\circ \times 1^\circ$ grid, (4) the daily tmp2m anomalies are computed by subtracting the climatology from the observed daily tmp2m, and (5) the forecasting target at each date and grid point is the average of tmp2m anomalies at day 15 to day 28. The climatology used in step (4) is the smoothed long-term average of tmp2m over 1990 - 2016 for each month-day combination and grid point. Specifically, for a given grid point, we compute the long-term average over 1990 - 2016, one for each month-day combination. Then the 365 values are smoothed using moving average with a window size of 31 days. 

\subsection{Evaluation Metrics}

We consider three metrics to evaluate the predictive performance of each forecasting model. Let $\y^*\in \R^n$ denotes a vector of ground truth temperature anomalies and $\hat{\y}\in \R^n$ are the corresponding predicted values. 

 \textit{(Uncentered) Anomaly Correlation Coefficient (ACC)} \cite{wilks2011statistical} is defined as 
    \begin{equation}
        \text{ACC}=\frac{\langle \hat{\y},\y^*\rangle}{\norm{\hat{\y}}_2\norm{\y^*}_2}~,
    \end{equation}
    where $\langle \hat{\y},\y^* \rangle$ denotes the inner product between the two vectors. Uncentered anomaly correlation is the only metric used in the Sub-Seasonal Climate Forecast Rodeo Competition \cite{raff2017sub, hwang2019improving}.

    
 \textit{Relative} $R^2$ is defined as 
\begin{align}
   \text{relative } R^2 &=1-\text{Relative MSE}\\
   &=
1-\frac{\sum_{i=1}^n(\y^*_i-\hat{\y}_i)^2}{\sum_{i=1}^n(\y^*_i-\bar{\y}_{\text{train}})^2}~, 
\end{align}
where $\bar{\y}_{\text{train}}$ is the long-term average of tmp2m anomalies at each date and grid point in the training set. Relative $R^2$ represents the relative skill against the best constant predictor, i.e., $\bar{\y}_{\text{train}}$. A model which achieves a positive relative $R^2$ is, at least, able to predict the sign of $y^*$ accurately and outperforms the climatology.


\textit{Root Mean Square Error (RMSE)} is defined as 
\begin{equation}
    \text{RMSE}= \sqrt{\frac{\sum_{i=1}^n(\y^*_i - \hat{\y}_i)^2}{n}}~.
\end{equation}
where $\y^*_i$ and $\hat{\y}_i$ are the $i$-th element in $\y$ and $\hat{\y}$ respectively.

Denote the ground truth temperature anomalies as $Y^*\in\R^{T\times G}$, where $T$ is the number of dates and $G$ is the number of grid points. The spatial predictive skill for a given date $t$ can be evaluated on $\y^*_t = Y^*[t, :]$, the $t$-th row in $Y^*$, where $\y^*_t\in\R^G$ is the ground truth for all grid points at date $t$ with the corresponding forecasts $\hat{\y}_t$. The temporal predictive skill for a grid point $g$ can be evaluated on $\y^*_g = Y^*[:, g]$, the $g$-th column in $Y^*$, similar to time series prediction evaluation.


\section{Subseasonal Experiment (SubX) Project}
\label{sec: subx}

\begin{table}[t]
\caption{Summary of GMAO-GEOS and NCEP-CFSv2.}
\label{table: data-subx}
\centering
\resizebox{0.7\columnwidth}{!}{
\begin{tabular}{c|c|c|c|c}
\hline
SubX Model & \makecell{Ensemble\\ Members} & \makecell{Initialization\\ Interval (day)} &\makecell{Hindcast\\ Range} & \makecell{Forecast\\ Range}\\
\hline
GMAO-GEOS & 4 & 5 & \makecell{01-01-1999 to \\ 12-31-2015} & \makecell{07-25-2017 to \\06-30-2020}\\
\hline
NCEP-CFSv2 & 4 & 1 & \makecell{01-01-1999 to \\ 12-31-2015} & \makecell{07-01-2017 to \\03-15-2019}\\
\hline
\end{tabular}}
\end{table}

\begin{table*}[t]
\caption{Description of climate variables and their data sources.}
\label{table: data}
\centering
\resizebox{\textwidth}{!}{
\begin{tabular}{c|c|c|c}
\hline
 Type &\makecell{Climate variable} &Description  & \makecell{Data Source} \\
\hline
\parbox[t]{2mm}{\multirow{7}{*}{\rotatebox[origin=c]{90}{Spatiotemporal}}}
&\makecell{tmp2m} &\makecell{Daily temperature at 2 meters} & 
\makecell{CPC Global Daily Temperature~\cite{fan2008global}}
\\
\hhline{~---}
&sm&\makecell{Monthly soil moisture} & \makecell{CPC Soil Moisture~\cite{soil_3}}\\
 \hhline{~---}
 & sst& \makecell{Daily sea surface  temperature}   & \makecell{Optimum Interpolation SST (OISST)~\cite{reynolds2007daily}}\\
 \hhline{~---}
& rhum& \makecell{Daily relative humidity \\near the surface  (sigma  level  0.995)}&\multirow{3}{*}{{\makecell{Atmospheric  Research\\ Reanalysi Dataset~\cite{reanalysis}}}}\\
\hhline{~--~}
&\makecell{slp}& \makecell{Daily pressure at sea level}  &                       \\ 
\hhline{~--~}
&\makecell{hgt10 \& hgt500} & \makecell{Daily geopotential height}    & \\
\hline
\parbox[t]{2mm}{\multirow{6}{*}{\rotatebox[origin=c]{90}{Temporal}}} &MEI.v2 & \makecell{Bimonthly multivariate ENSO index}&  \makecell{NOAA ESRL MEI.v2~\cite{zhang2019towards}}\\ 
\hhline{~---}
& \makecell{MJO phase \& amplitude}& \makecell{Madden-Julian Oscillation index}&     \makecell{Australian Government BoM  \cite{MJO}}\\
\hhline{~---}
& \makecell{Ni\~no 1+2, 3, 3.4, 4}& \makecell{Weekly Ni\~no index}&     \makecell{NOAA National Weather Service, CPC~\cite{reynolds2007daily}}\\
\hhline{~---}
& \makecell{NAO}& \makecell{Daily North Atlantic Oscillation index}&    \makecell{NOAA National Weather Service, CPC~\cite{NAO_2}}\\
\hhline{~---}
& \makecell{SSW}& \makecell{Sudden  Stratospheric  Warming  index \\(The zonal  mean  winds  at  60N  at  10hPa)}&   \makecell{Modern-Era Retrospective Analysis \\for Research and Applications v2 \cite{gelaro2017modern}}\\
\hline
\end{tabular}}
\end{table*}


The Subseasonal Experiment (SubX) is a project that provides sub-seasonal forecasts from multiple global forecast models.  Data are publicly available through the International Research Institute for Climate and Society (IRI) Data Library at Columbia University. A detailed description of the SubX project and the contributing models can be found in \cite{subX}. The SubX project has two predictive periods: hindcast and forecast. A hindcast period represents the time when a dynamic model re-forecasts historical events, which can help climate scientists develop and test new models to improve forecasting. Hindcasts in the SubX project occurred during January 1999 to December 2015. In contrast, a forecast period has real-time predictions generated from dynamic models. The real-time forecast period in the SubX project starts from July 2017. We evaluate the predictive skills of the SubX models over their forecast periods. 


In this paper, we focus on two SubX models, NCEP-Climate Forecast System version 2 (CFSv2) \cite{saha2014ncep} and NASA-Global Modeling and Assimilation (GMAO) version 2 of the Goddard Earth Observing System (GEOS) model \cite{reichle2014observation}. NCEP-CFSv2 is the operational seasonal prediction model currently used by the U.S. Climate Prediction Center. The NCEP-CFSv2 forecasts are initialized daily and include four ensemble members. In order to ensure our results are not unique to a single dynamical model, we also analyze output from the GMAO-GEOS model, which is developed to support NASA's earth science research. GMAO-GEOS is also a fully coupled atmosphere–ocean–land–sea ice model, with forecasts initialized every five days and includes four ensemble members. We selected these two models since they have the highest initialization frequency in the SubX project. Other SubX models were initialized less frequently. Besides, our code base and the ground truth dataset are fairly flexible, which can easily be extended to evaluate other SubX models.


Further information of the two SubX models are presented in Table \ref{table: data-subx}. For both SubX models, the average of four ensemble members' outputs are taken as the forecasts. All forecasts include daily values for 45 days beyond the initialization date. The weeks 3 \& 4 outlooks are computed by averaging the forecasts 15 to 28 days beyond each initialization date and subtracting the corresponding climatology computed from the model's hindcast period.


\section{Machine Learning-based SSF Modeling}
\label{sec: ml}
\noindent\textbf{Notation.} Let $Y\in \R^{T\times G}$ denote the targeted weeks 3 \& 4 temperature anomalies over $T$ dates and $G$ grid points. $\y_t\in \R^G$ is the $t$-th row in $Y$, denoting the temperature anomalies over all grid points $G$ at date $t$. $X\in \R^{T\times p}$ denotes the $p$-dimension covariates for $T$ dates. $\x_{t}\in \R^p$ (the $t$-th row in $X$) is the covariate at date $t$.


\subsection{Machine Learning Models}

In this paper, we focus on state-of-the-art machine learning models which have been shown to work effectively for sub-seasonal climate forecasting \cite{heli21,hwang2019improving}.

\textit{AutoKNN} \cite{hwang2019improving}. An auto-regressive model only uses features from historical temperature anomalies, which selects lagged measurements with a multitask $k$-nearest neighbor criterion. For a given date $t$, the algorithm chooses the temperature anomalies from 20 historical dates with the highest similarity and 29 days, 58 days, and 1 year prior to $t$ as features. Specifically, the similarity between two dates $t_1$ and $t_2$ is defined as $\text{sim}_{(t_1, t_2)}=\frac{1}{M}\sum_{m=0}^{M-1} \cos(\y_{t_1-l-m},\y_{t_2-l-m})$, where $\cos(\y_{t_1-l-m},\y_{t_2-l-m})$ is the cosine similarity between the temperature anomalies at $l-m$ days before $t_1$ and $t_2$. Following the settings in \cite{hwang2019improving}, we use $M=60$, the length of the considered historical sequences prior to each date, with the lag $l=365$. At each grid point, we fit a weighted local linear regression model, where the weight is one over the variance of the temperature anomalies at the corresponding date.



\textit{Multitask Lasso} \cite{lasso,jalali2013dirty}. A multitask regularized linear regression model. 
By assuming $\y_{t} = \x_t^T\Theta^* + \epsilon$, where $\epsilon \in \R^G$ is a Gaussian noise and $\Theta^*\in \R^{p\times G}$ is the coefficient matrix for all locations, the parameter $\Theta^*$ is estimated by
\begin{equation}
    \hat{\Theta} = \argmin_{\Theta\in \R^{p\times G}}\frac{1}{2T} \|Y - X\Theta\|_2^2 + \lambda \norm{\Theta}_{2,1}
\end{equation}
with $||\Theta||_{2,1} = \sum_{i} (\sum_{j} \Theta_{ij}^2)^{1/2}$ and $\lambda$ being the penalty parameter.


\textit{Gradient boosted trees (XGBoost)} \cite{friedman2001greedy,xgboost}. A functional gradient boosting algorithm, of which the weak learners are regression trees. The algorithm combines multiple weak learners into a stronger learner in an iterative manner. At each iteration, a new weak learner is created to correct the previous prediction and optimize the loss function along with regularization. We build one XGBoost model for each location, and the hyper-parameters are selected jointly based on the performance over all locations.


\begin{figure}[t]
    \centering
    \includegraphics[width=0.55\textwidth]{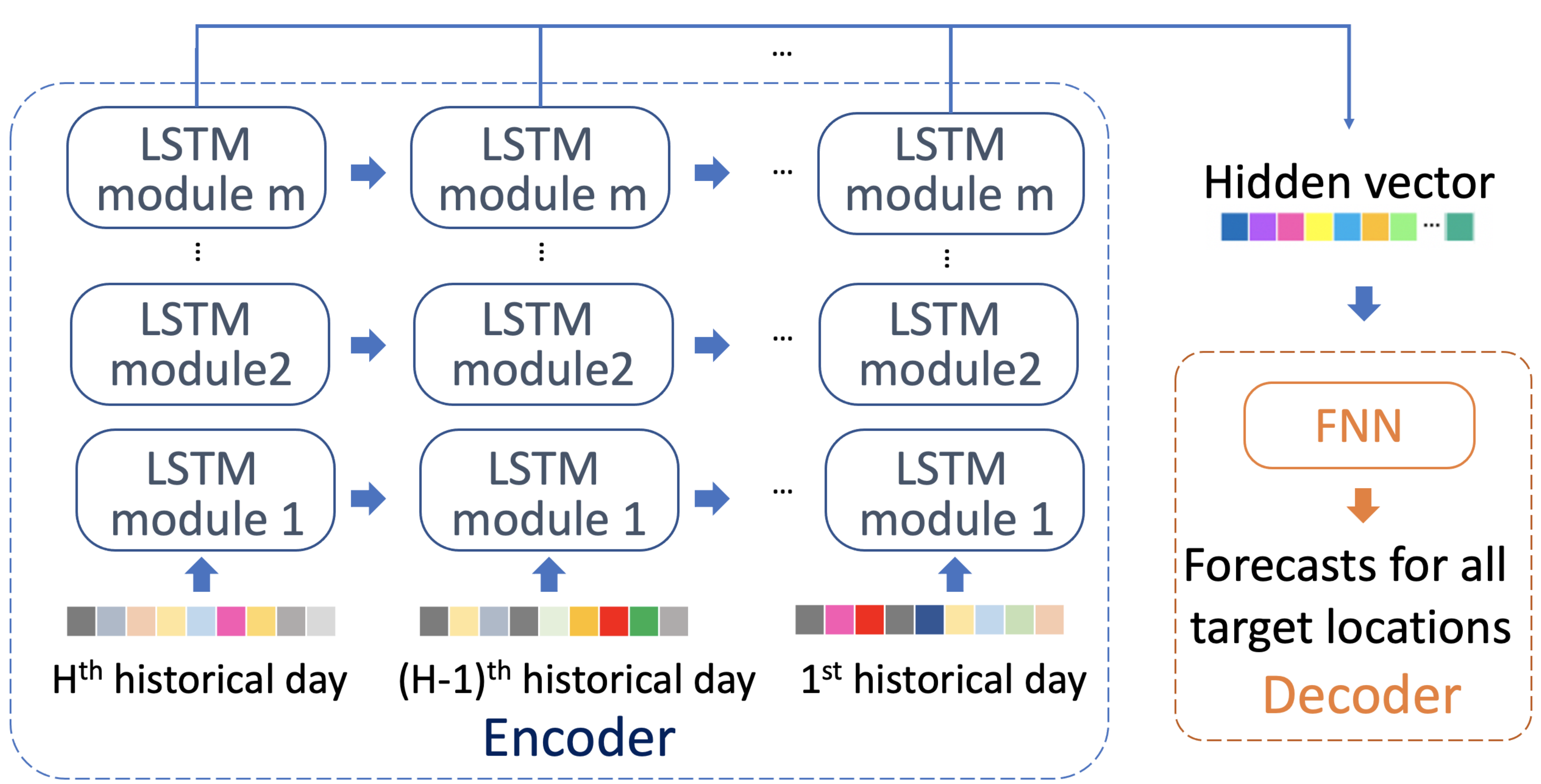}
    \caption{Architectures of the Encoder-FNN that has a
few LSTM layers as the Encoder, and two fully connected layers as the Decoder \cite{heli21}.}
     \label{fig:model_architecture}
\end{figure}

\textit{Encoder-FNN}\cite{heli21}. A deep learning model designed for SSF over the contiguous U.S., of which the architecture is shown in Figure \ref{fig:model_architecture}. The model input is a historical sequence of the features shared by all locations and is fed into an LSTM encoder recurrently. The output of each step in the sequence are combined and jointly sent to the decoder, which is a two-layer fully-connected neural network with ReLU activation. The outputs of the decoder are the predicted tmp2m anomalies over all grid points. Note that, besides standard hyper-parameters like layer size, number of layers, and dropout rate, the length of the sequence is also a hyper-parameter. The final forecast is the average of 20 independent runs. 


\subsection{Covariates for ML models}

The feature set for the ML models contains the following climate variables.  
Spatially over the contiguous U.S. we consider (1) 2m temperature (tmp2m), which is also the source data for the ground truth dataset; (2) soil moisture, which influences temperature and precipitation through its impact on surface fluxes of heat and moisture \cite{koster2011}; and (3) four climate variables - geopotential height (ght) at 10mb and 500mb, sea level pressure (slp) and relative humidity (rhum) - from the reanalysis dataset, which capture variations in the northern hemisphere polar vortex and persistent variations in the large-scale atmospheric circulation. We also obtain sea surface temperature (sst) over the Pacific Ocean, from latitudes 20S to 65N and longitudes 120E to 90W, and the Atlantic Ocean, from latitudes 20S to 50N and longitudes 20W to 90W. Variations in sst have been linked to enhanced sub-seasonal predictability over the U.S. \cite{delsole2017}. 

In addition, we include nine climate indices that describe the state of the climate system or are related to different climate phenomena, such as El Ni\~no/Southern Oscillation (ENSO). Multivariate ENSO index (MEI.v2) and Ni\~no indices are included for monitoring  El Ni\~no and La Ni\~na events \cite{delsole2017,stan2017}. The amplitude and phase of Madden-Julian Oscillation (MJO) are considered since the MJO has dramatic impacts in the mid-latitudes and is a strong contributor to various extreme events in the U.S. \cite{Waliser2005}. North Atlantic Oscillation index (NAO) is considered since variations in the NAO drive changes in temperature and precipitation over the U.S. and western Europe \cite{stan2017}. Sudden  Stratospheric  Warming  index (SSW) is included to capture the variations in the strength of the polar vortex, which are associated with extreme cold air outbreaks in mid-latitude U.S., Europe, and Asia \cite{butler2015}.

\subsection{Data Preprocessing}

For all ML models except AutoKNN, we consider two types of climate variables, namely spatiotemporal climate variables and temporal climate variables. For each spatiotemporal variable, we flatten the values at all grid points for each date and compute the top 10 principal components (PCs) as features. For example, if $X^{\text{sm}}\in \R^{T\times G}$ denotes the soil moisture for $T$ dates in training set (1990-2016) and all $G$ spatial grid points over the contiguous U.S., we compute the PC loadings using $X^{\text{sm}}$ and extract the top 10 PCs to get the feature matrix $X^{\text{sm}}_{\text{pc}}\in \R^{T \times 10}$. The extracted PCs are then normalized by z-scoring for each month-day combination separately. The temporal variables and the PC-based features of all spatiotemporal climate variables jointly form the feature set for each date. For XGBoost and Lasso, the covariates are the feature values two weeks lagging from the forecasting period. For example, if the forecasting period is Jan, 15 - Jan, 28 in 2019, the covariates are the features on Jan 1, 2019. For Encoder FNN, the features of a historical sequence are treated as the model input for each date. The historical sequence is constructed similarly to the features of Encoder FNN in \cite{heli21} (Figure 7(a)). AutoKNN takes only the historical tmp2m anomalies as the covariate.




\subsection{Experimental Setup}
Since the relationships between the covariates and target variables vary at different times of the year, test sets are created for each month from July 2017 to Jun 2020 and separate predictive models are trained accordingly. 
Since an individual ML model is built for each month of the year, the best hyper-parameters of each type of ML models are selected on a monthly basis. To do so, for each month of the year, we construct five validation sets containing data from the same month between 2012 and 2016, and the corresponding training sets consist of 10 years of data prior to each validation set. The best hyper-parameters are determined by the average performance over the five validations sets. We thus have twelve sets of the best hyper-parameters corresponding to each month of the year. Once the best hyper-parameters are selected, we use 28 years of data prior to a given test set to train the corresponding ML forecasting model.



\begin{figure}[ht]
\centering
\subfigure[The empirical cdf and QQ plot for spatial relative $R^2$]{\includegraphics[width=0.48\columnwidth]{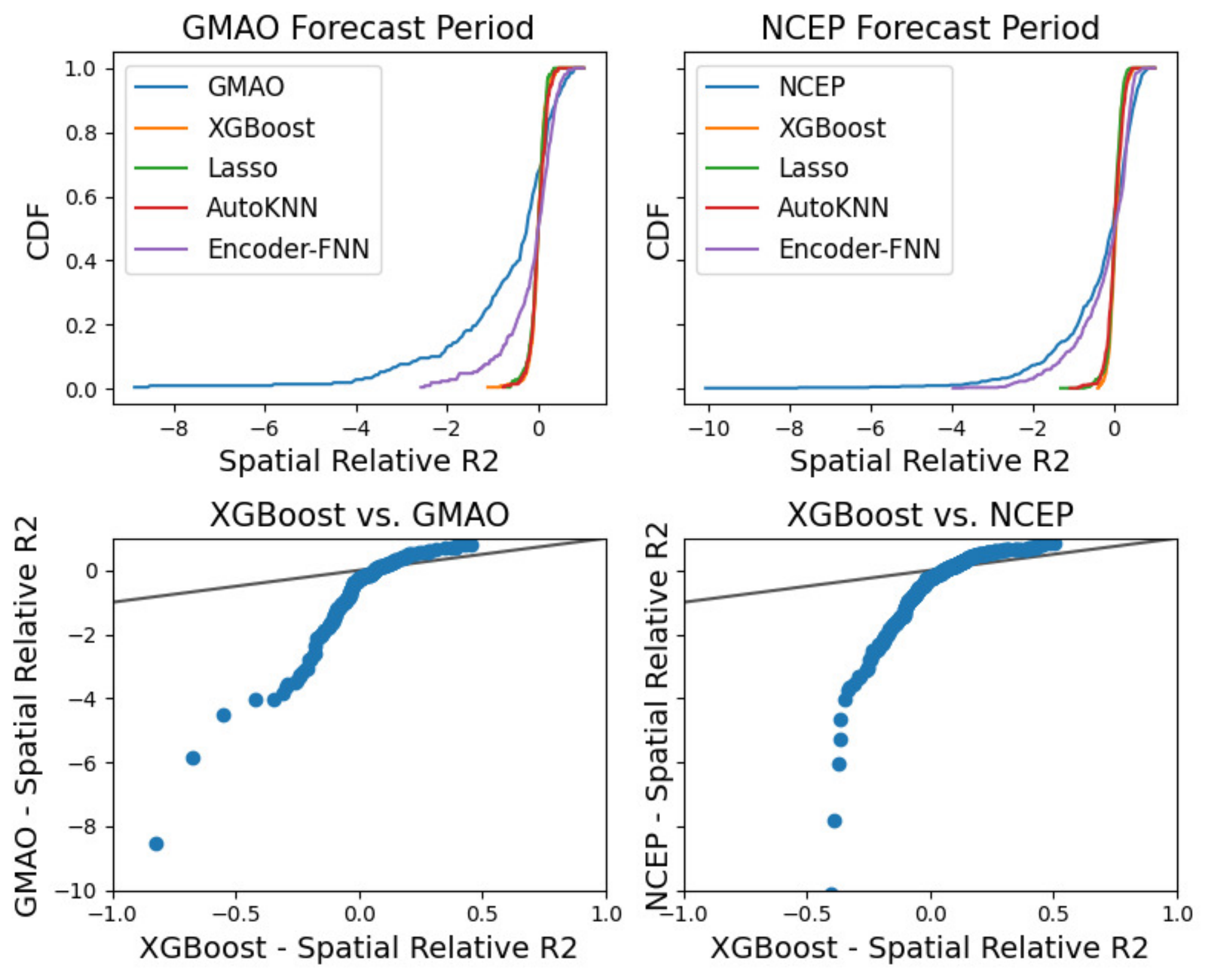}}
\subfigure[The empirical cdf and QQ plot for spatial ACC]{\includegraphics[width=0.48\columnwidth]{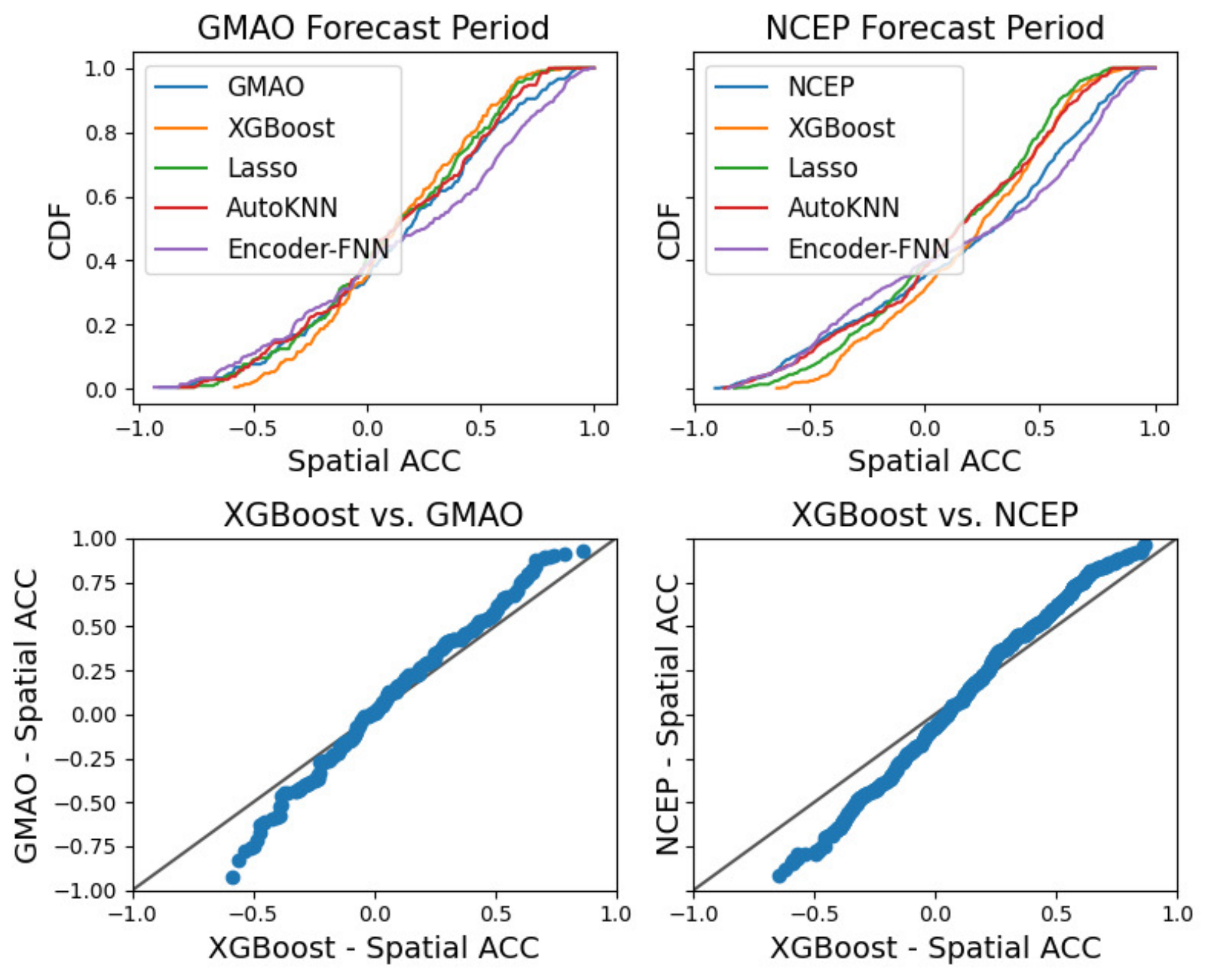}}
\caption{(a) The empirical cumulative distribution function (cdf) of spatial relative $R^2$ (top) of all methods and the quantile-quantile (QQ) plot of relative $R^2$ (bottom) between XGBoost and GMAO-GEOS (left) or NCEP-CFSv2 (right). XGBoost, Lasso and AutoKNN all have most spatial relative $R^2$ close to or above 0, while the SubX models and Encoder-FNN have relative $R^2$ much smaller than -1. (b) The cdf and QQ plot of spatial ACC. Despite the similarity of the cdf curves, the ML models (yellow, green, and red) are in general below the blue curve (the SubX model) when the spatial ACCs are negative, which indicates that the ML models are less likely to have extremely negative predictive skills compared to the SubX models.}
\label{fig:all_results}
\end{figure}

\section{Experimental Results}
\label{sec: exp-res}

\begin{figure*}[t]
\centering
\subfigure[Temporal ACC]{\includegraphics[width=0.85\textwidth]{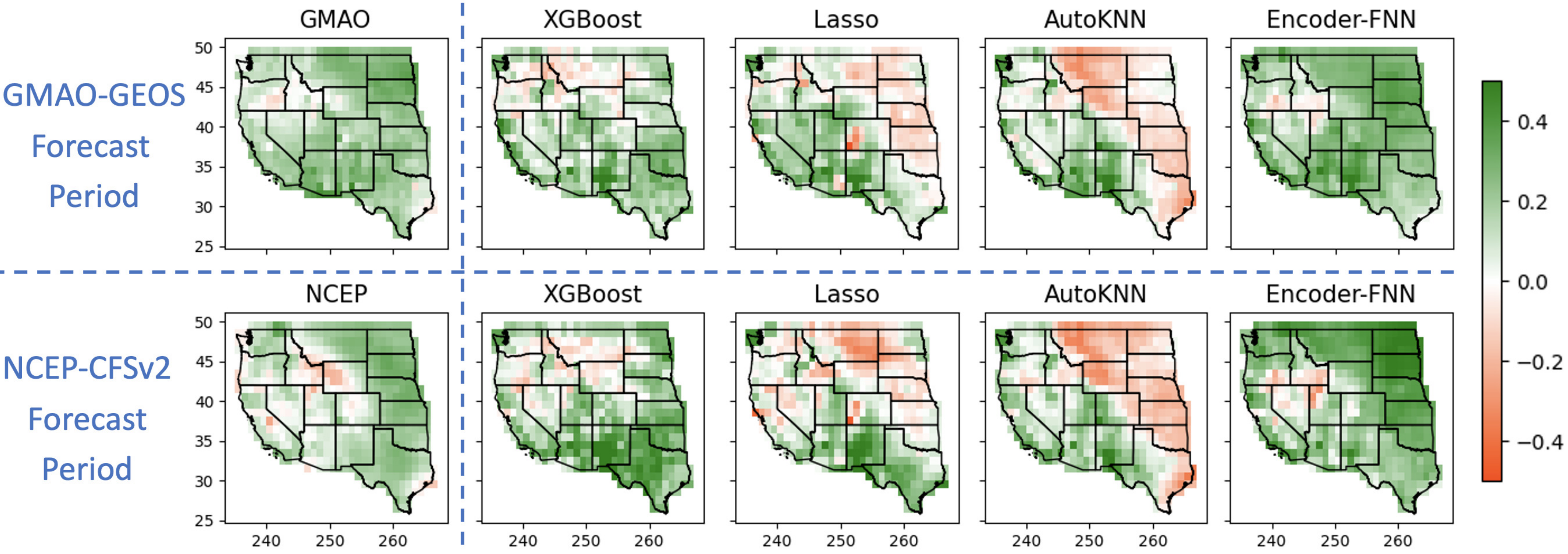}}
\subfigure[Temporal Relative $R^2$]{\includegraphics[width=0.85\textwidth]{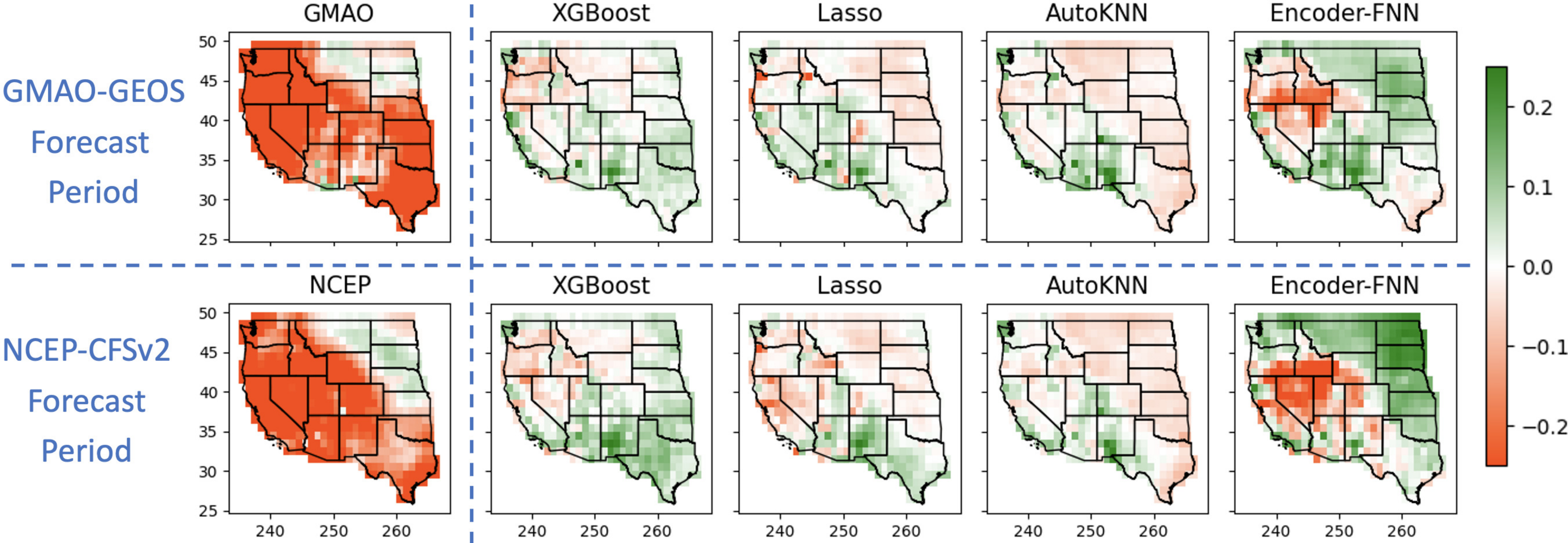}}
\caption{(a) Temporal ACC  and (b) temporal relative $R^2$ of the SubX models (leftmost columns) and the ML models. For both metrics, values closer to 1 (green) indicate more accurate predictions. Overall the SubX models achieve positive temporal ACC for most spatial locations while performing poorly if considering temporal relative $R^2$. Among all the ML models, XGBoost and Encoder-FNN are the two best models regarding both predictive skills, and substantially outperform the SubX models over most spatial locations, especially compared to NCEP-CFSv2.}
\label{fig:all_results_map}
\end{figure*}

In this section, we compare the predictive skill of the four ML models and the two SubX models on the forecast period from 2017 to 2020. A comprehensive analysis is conducted for the experimental results, which reveals possible directions for further improvement of the ML models for SSF. Besides, we explore the potential of advancing SSF by combining the ML models and the SubX forecasts.

\subsection{Forecast Period Evaluation}
Since the GMAO-GEOS and NCEP-CFSv2 models have different forecast periods and temporal resolutions, all results are evaluated at their respective forecast periods and resolutions. We first present the empirical cumulative distribution function (cdf) of spatial relative $R^2$ for all methods over the forecast periods of GMAO-GEOS and NCEP-CFSv2 in Figure \ref{fig:all_results}(a). It is shown that ML models such as XGBoost, Lasso, and AutoKNN are capable of generating forecasts with positive or smaller negative relative $R^2$, while the SubX models and the Encoder-FNN model commonly stay in the negative relative $R^2$ zone. On the other hand, considering the positive side of the cdf plot, the SubX model and Encoder-FNN are able to achieve relative $R^2$ close to 1 in some cases, whereas the cdf of other ML models reach 1 when the relative $R^2$ are comparatively small. The quantile-quantile (QQ) plot of spatial relative $R^2$ in Figure \ref{fig:all_results}(a) shows that the relative $R^2$ can be much smaller than -1, indicating the SubX models can make predictions with a large deviation from the ground truth. A similar set of plots for spatial ACC is presented in Figure \ref{fig:all_results}(b). Despite the similarities in the cdf across models, a closer inspection shows the cdf of the ML models (yellow, green, and red curves) are generally below the cdf of the SubX models (blue curve) for spatial ACC between [-1, 0]. The QQ plot of the spatial ACC between XGBoost and SubX models supports the observation, where all the points are below the diagonal line when the spatial ACC of XGBoost is between [-1, 0]. For the positive side of the spatial ACC for XGBoost, most points are close to or slightly above the diagonal line. To summarize, at a given date, the SubX models are more likely to have spatial ACC close to the extreme values (-1 or +1), while ML models, such as XGBoost, are more conservative and are able to avoid extreme negative ACC. 

The temporal ACC and temporal relative $R^2$ over the western U.S. are illustrated in Figure \ref{fig:all_results_map}(a) and (b) respectively. Similar to spatial results, the SubX models achieve positive temporal ACC for most spatial locations while performing poorly with respect to temporal relative $R^2$. Among all ML models, XGBoost and Encoder-FNN are the best two considering temporal predictive skills and substantially outperform the SubX models for most spatial locations, especially compared to NCEP-CFSv2. Spatially, the central area, including the states of North Dakota, South Dakota, Montana, Wyoming, Kansas, and Oklahoma, are the areas where the temperature fluctuations are more drastic compared to the coastal states. Therefore, linear model like Lasso and non-parametric model like AutoKNN tend to perform worse in such regions, while more complicated nonlinear models like XGBoost and Encoder-FNN perform relatively better. Additionally, the SubX models have negative temporal relative $R^2$ and positive temporal ACC for the coastal area, which implies the SubX models may predict incorrect magnitudes despite their relatively accurate prediction of the temporal patterns.

\begin{figure}[t]
\centering
\includegraphics[width=0.45\textwidth]{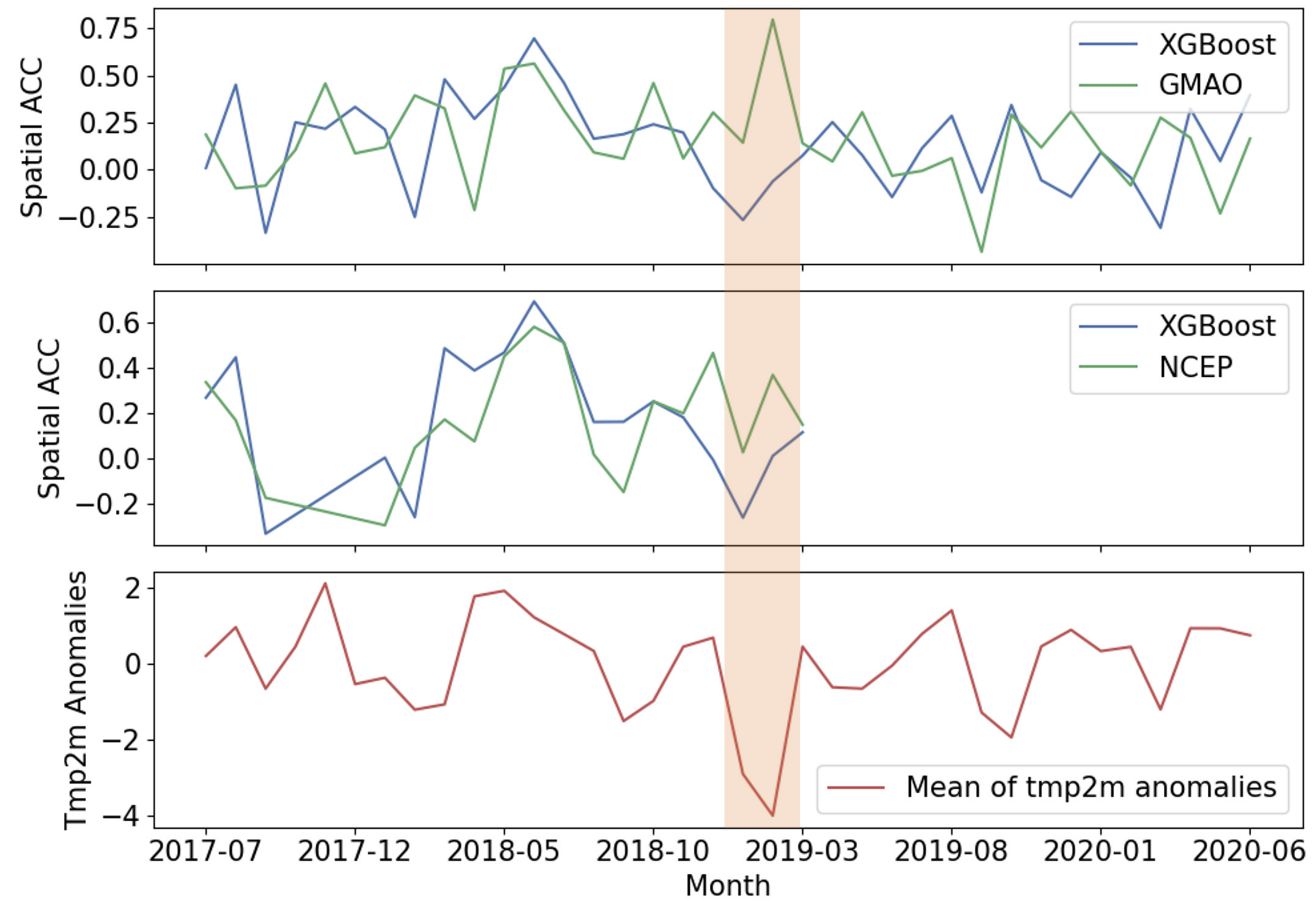}
\caption{The monthly average spatial ACC of XGBoost and the SubX models (top 2) during the respective forecast periods and the mean of tmp2m anomalies over the western U.S. (bottom). Most of the time, XGBoost achieves competitive or even higher spatial ACC compared to the SubX models. The only exception, that both SubX models outperform XGBoost, happens from Dec. 2018 to Feb. 2019 (highlighted in orange) when a cold wave affected the U.S. leading to extreme low average tmp2m anomalies.} 
\label{fig:acc-month}
\end{figure}

\begin{figure}[htb]
    \centering
    \includegraphics[width=0.45\columnwidth]{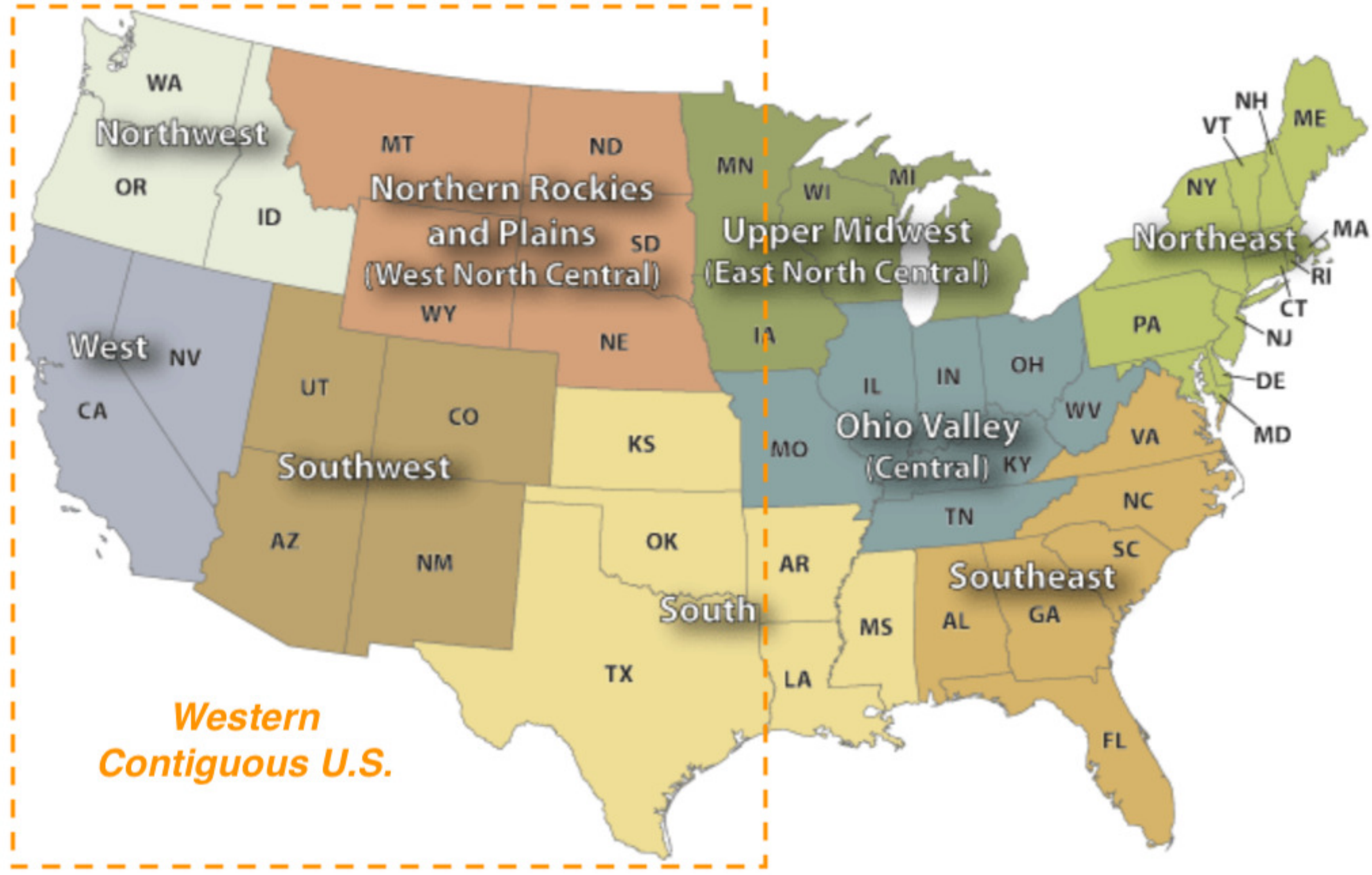}
    \caption{The nine climatically consistent regions identified by National Centers for Environmental Information scientists \cite{karl1984regional, climate_region}. The western contiguous U.S. (orange rectangular) covers five regions with 20 states.}
     \label{fig:climate-region}
\end{figure}

\begin{figure}[t]
\centering
\subfigure[Spatial RMSE vs. std of tmp2m anomalies 
]{\includegraphics[width=0.48\textwidth]{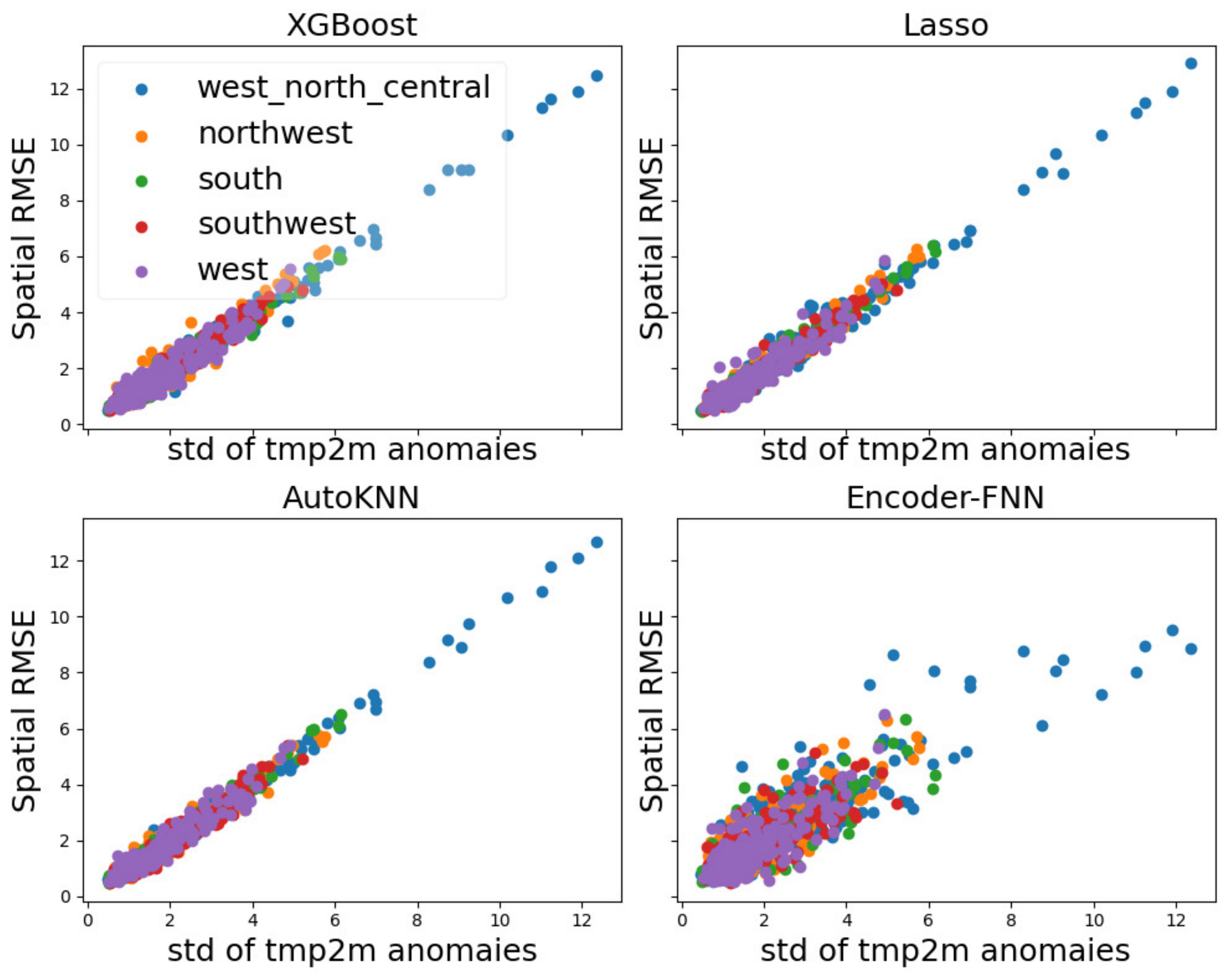}}
\subfigure[Spatial RMSE vs average tmp2m anomalies 
]{\includegraphics[width=0.48\textwidth]{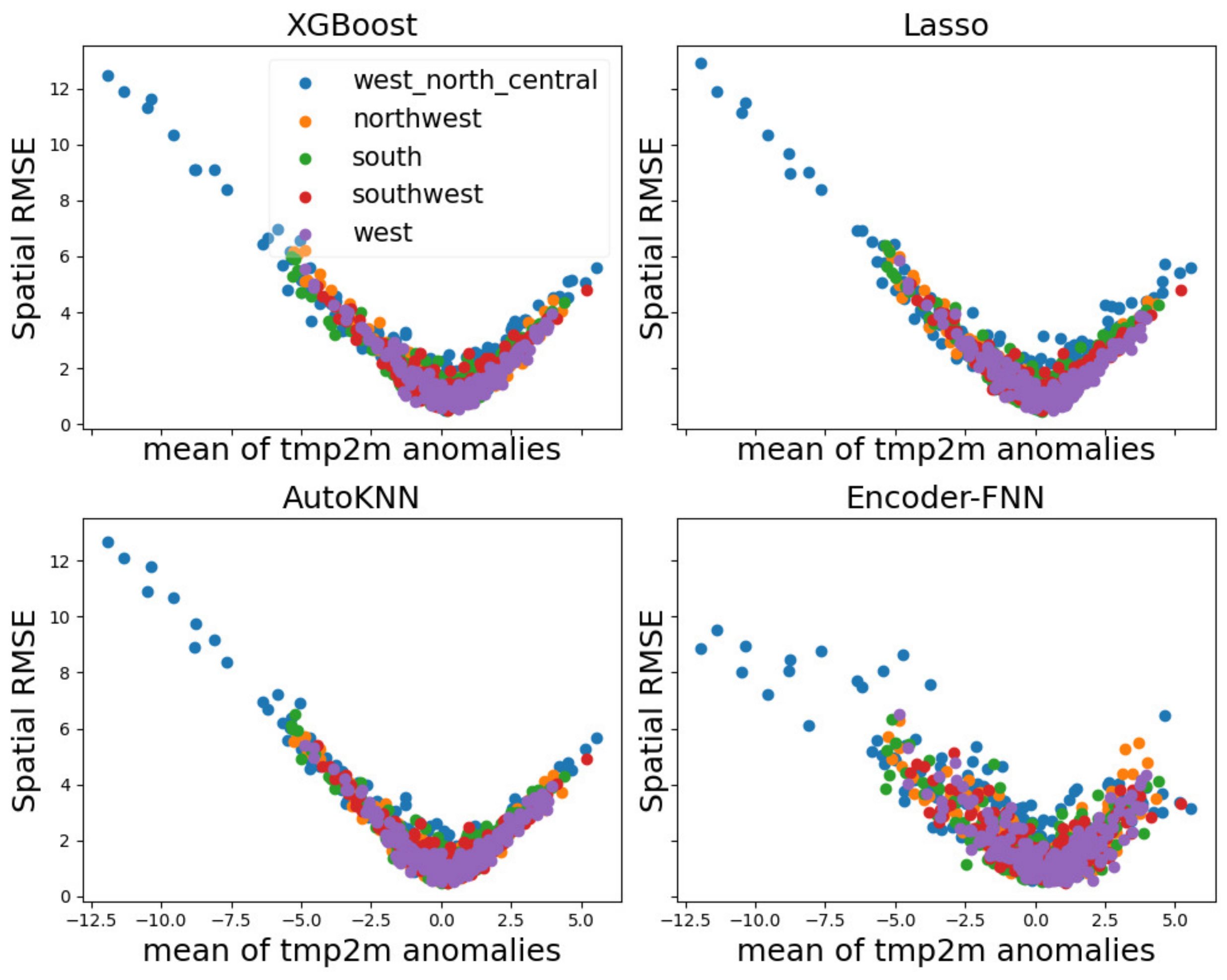}}
\caption{Spatial RMSE versus (a) the standard deviation of tmp2m anomalies over dates and regions and (b) the average tmp2m anomalies over the dates and regions. The high spatial RMSE appears for samples having large standard deviation or extreme negative average of tmp2m anomalies, which indicates that the west-north-central region are hard to predict.
}
\label{fig: rmse_vs_mean}
\end{figure}

\subsection{Machine Learning and Extreme Weather Events}
Given that SSF is a challenging problem, it is natural to investigate under which circumstance(s) the ML models fail to provide accurate forecasts. The average spatial ACC of the XGBoost models and the SubX models for each month during the forecast periods are shown in Figure \ref{fig:acc-month}. For most months, XGBoost is either competitive or achieves higher spatial ACC compared to the SubX models. The exceptions occur in December 2018 and first two months of 2019, when the January–February 2019 North American cold wave impacted the United States. 
The cold wave brought the coldest temperatures in over 20 years to most locations \cite{cold-wave}. The temperature anomalies reached $-15^{\circ}C$ and beyond in the \textit{central U.S.} Extreme weather events are hard to predict since there is a lack of enough training data for such events. However, the dynamical models are reasonably successful in predicting the extreme cold temperatures, since they follow the physics. For example, the cold wave followed a sudden stratospheric warming event, which have been shown to increase predictability of these extreme events \cite{domeisen2020stratospheric}.

\begin{figure}[t]
\centering
\subfigure[Ground truth and forecasts on Mar. 12, 2018]{\includegraphics[width=0.65\columnwidth]{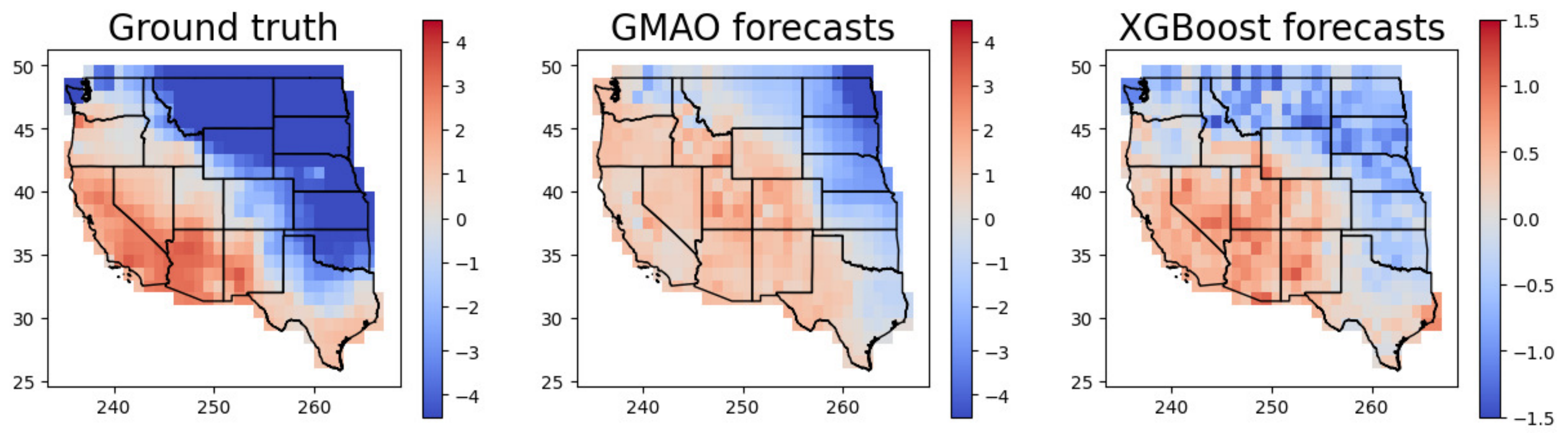}}
\subfigure[Ground truth and forecasts on Jan. 6, 2020]{\includegraphics[width=0.65\columnwidth]{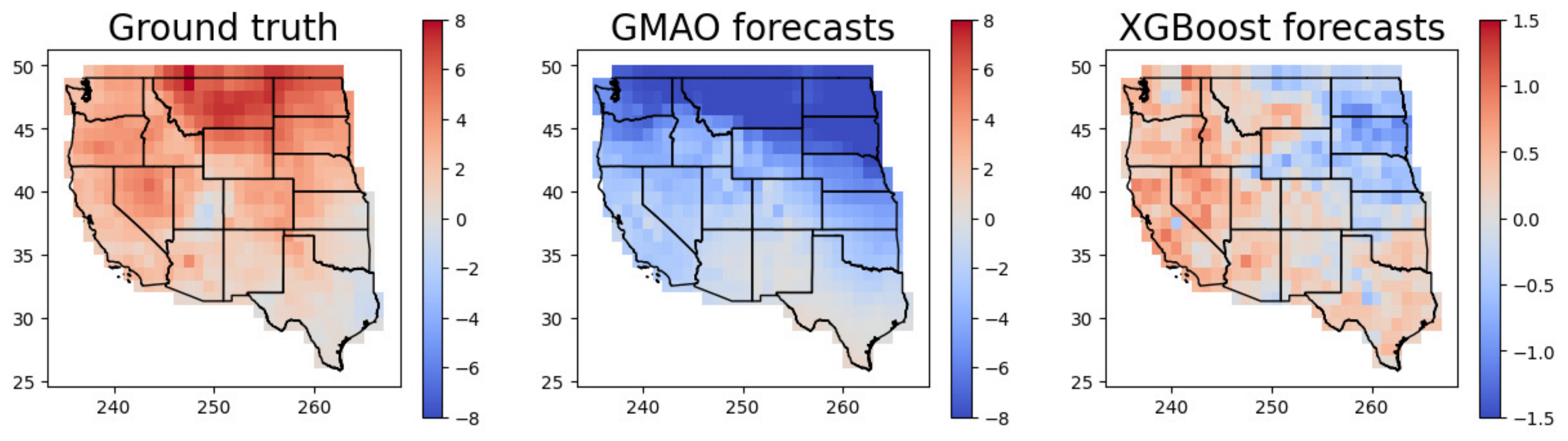}}
\caption{Comparison between the ground truth and forecasts made by GMAO-GEOS and XGBoost at two selected dates. (a) On March 12, 2018, both XGBoost and GMAO-GEOS successfully predict the spatial pattern of ground truth data (red in the southwest and blue in the northeast). However, predicted values from XGBoost are much smaller than GMAO-GEOS forecasts as XGBoost is more conservative on the magnitude. (b) On Jan 6, 2020, the ground truth has positive tmp2m anomalies (red) for most locations, while GMAO-GEOS mistakenly makes extreme negative forecasts (dark blue). }
\label{fig:example}
\end{figure}

The value of spatial ACC is not affected by the scale of the response. Therefore, we analyze the predictive performance regarding RMSE for the ML models. We first separate all grid points in the western U.S. into five climatically consistent regions \cite{karl1984regional}, i.e., northwest, west, west-north-central, southwest, and south (Figure \ref{fig:climate-region}). To represent the spatial variance of tmp2m anomalies at each forecasting date and each region, we approximately compute the standard deviation (std) of tmp2m anomalies at each date and each region as $\sqrt{\sum_{i=1}^{n_r}\frac{y_i^2}{n_r}}$, where $n_r$ is the number of grid points for a given region at one date. As shown in Figure \ref{fig: rmse_vs_mean}(a), the RMSE from all four ML models at a given date and region is strongly correlated to the std of tmp2m anomalies, which implies the dates and regions with high variance are difficult to predict. Figure \ref{fig: rmse_vs_mean}(b) illustrates the average of tmp2m (with sign, unlike Figure \ref{fig: rmse_vs_mean}(a)) for each date and region versus the predictive RMSE, which further demonstrates that extreme events are the samples with negative bias and large variance during the forecast period. Besides, the distribution of different regions in Figure \ref{fig: rmse_vs_mean} implies that the spatial variance is, in general, lower for coastal regions compared to inland regions. For instance, west-north-central region can experience extremely cold winter temperatures when the polar jet stream sinks down into the mid-latitudes and brings with it the coldest polar air.

This analysis illustrates the difficulty of modeling extreme weather events using a single ML model, not only because of the inadequate samples, but also due to the intense temperature fluctuations caused by such events. Therefore, it is necessary to utilize separate modeling techniques for weather extremes or regions with drastic fluctuations in tmp2m anomalies, to achieve more accurate predictions. Ideally, if weather extremes can be detected ahead of time, we can choose not to trust the ML forecasts for a certain time period and turn to the forecasting models specifically designed for extreme conditions.

\subsection{Enhancing ML models with SubX Forecasts}

To demonstrate the strengths and limitations of the SubX and the ML model forecasts, we present forecasts of two days as anecdotal evidence in Figure \ref{fig:example}. The first example (Figure \ref{fig:example}(a)) shows that, on Mar. 12, 2018, both GMAO-GEOS and XGBoost have successfully reproduced the spatial pattern of the ground truth. As a result, GMAO-GEOS and XGBoost obtain good spatial ACC. However, the predicted scale from GMAO-GEOS is much larger than XGBoost and is closer to the scale of the ground truth. The second example is the forecasting results on Jan 6, 2020, when the SubX forecasts fail badly. As shown in Figure \ref{fig:example}(b), while the ground truth is that all the locations over the western U.S. have positive tmp2m anomalies with the largest values around 8$^{\circ}C$, GMAO-GEOS predicts all negative tmp2m anomalies with the lowest values close to $-8^{\circ}C$. Meanwhile, XGBoost partially predicts the correct spatial pattern but with conservative values in the range of $[-1.5 ^{\circ}C, 1.5 ^{\circ}C]$, which are much smaller than the magnitudes of the ground truth. These two examples demonstrate that the SubX models have certain advantage on matching the magnitude of the tmp2m anomalies, while the ML models are more conservative and provide predicted values with smaller amplitude. On the flip side, in situations where the SubX models does not predict the spatial pattern correctly, the forecasts can be wrong by a large amount.

\begin{figure}[ht]
\centering
\subfigure[Temporal ACC]{\includegraphics[width=0.48\textwidth]{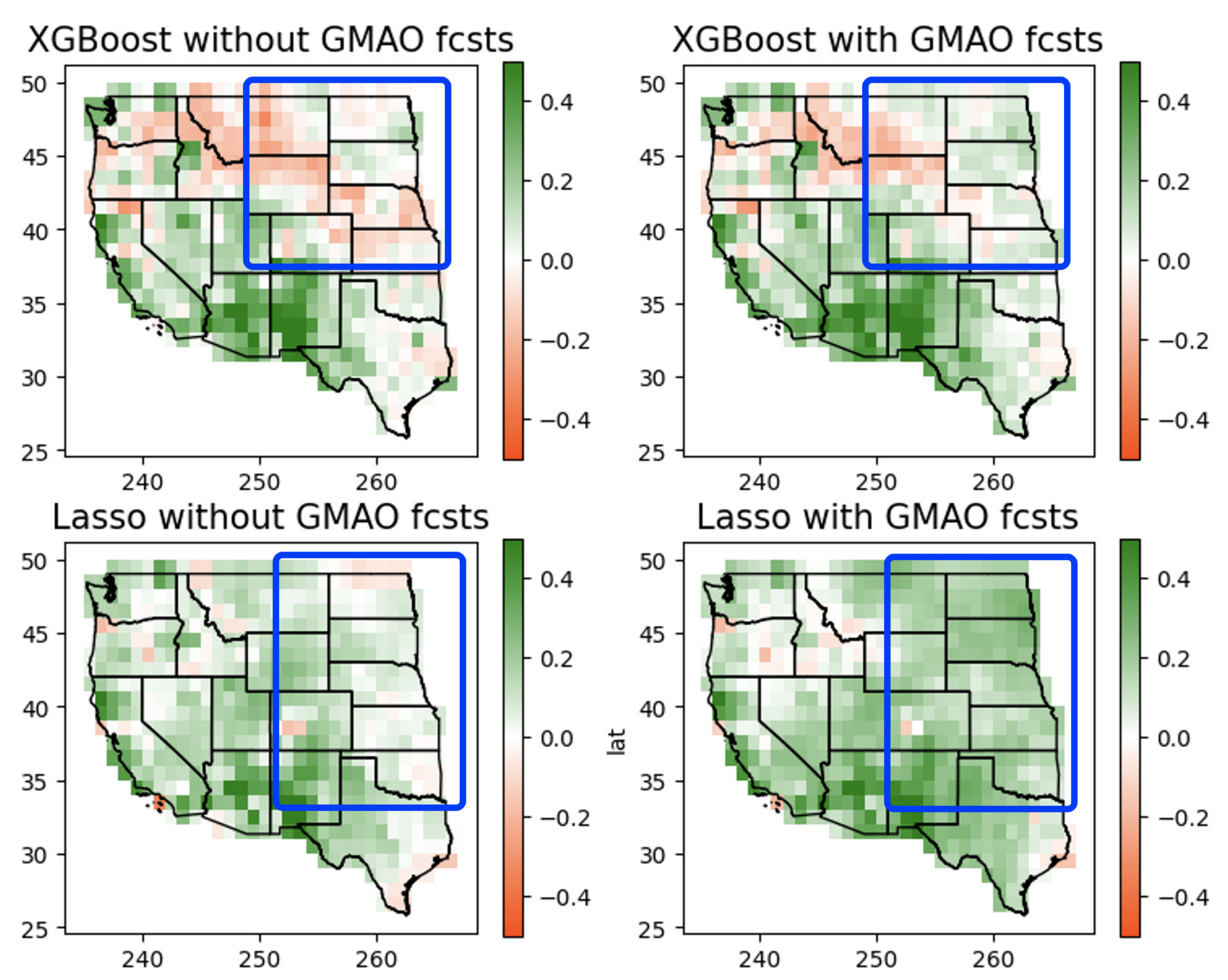}}
\subfigure[Temporal relative $R^2$]{\includegraphics[width=0.48\textwidth]{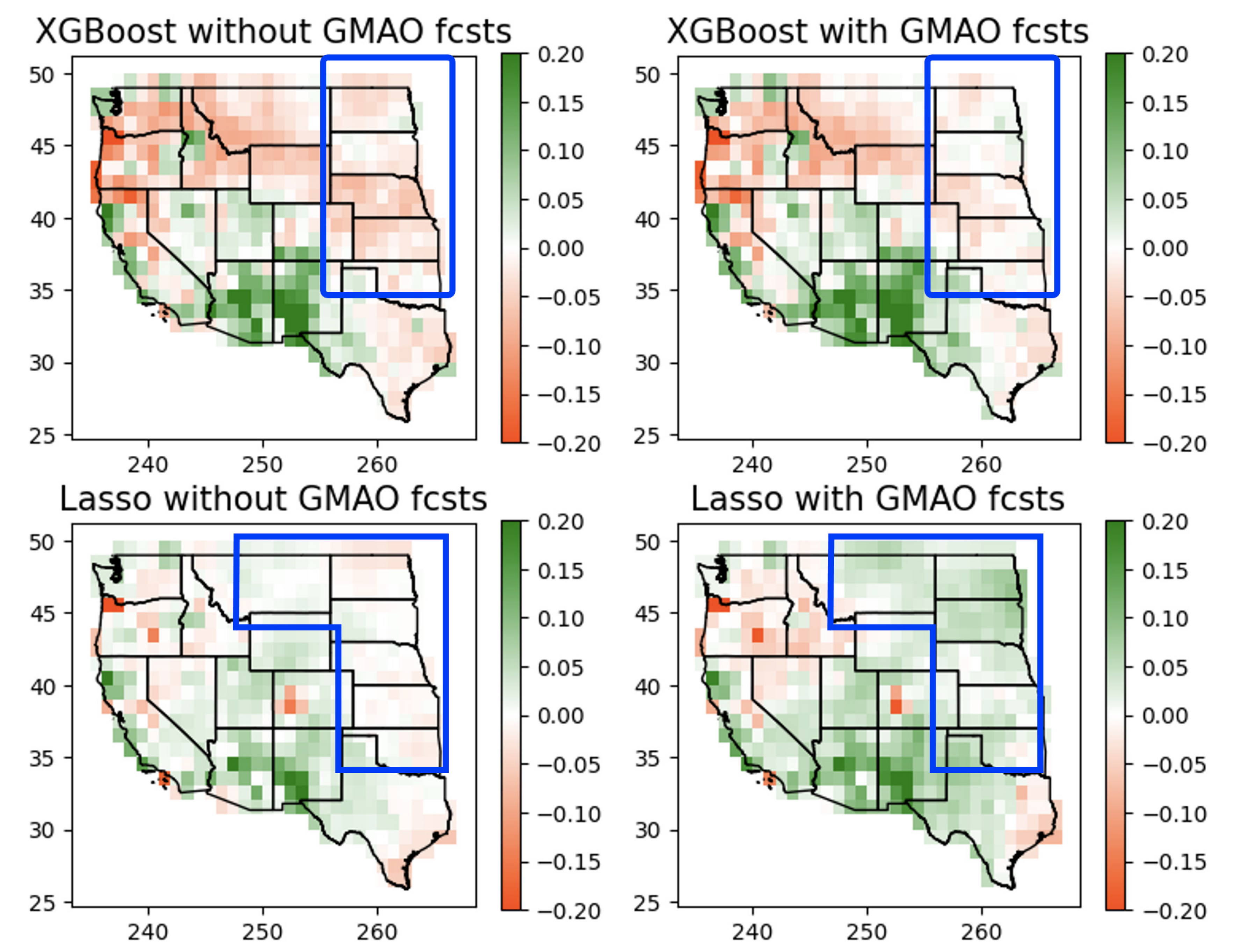}}
\caption{The temporal ACC and relative $R^2$ of XGBoost and Lasso with and without GMAO-GEOS forecasts as features. Including GMAO-GEOS forecasts in the feature sets evidently improves the forecasting performance, especially for the central U.S. (top right corner, marked by blue frames).}
\label{fig: ml_subx_gmao}
\end{figure}

Acknowledging the advantages of both types of models, we explore a suitable combination of the ML models and the SubX forecasts. More specifically, we investigate whether including SubX forecasts in the feature set of the ML models can enhance the predictive skill of the ML models. Since the hindcast periods of the SubX models are $\sim$10 years shorter than the temporal range of the training data for the ML models, and the temporal resolution of SubX models is also relatively lower, incorporating the SubX forecasts significantly reduces the sample size. To compare the performance fairly, we first train a ML model using the samples that are available during the hindcast periods and then compare it with the ML model that uses SubX forecasts as features, this guarantees both models are trained with exactly the same sample size. Note, for Multitask Lasso, features are originally shared for all locations. To incorporate SubX forecasts, we have to build one Lasso model for each location but the hyper-parameter is jointly selected based on the performance for all locations.

\begin{table}[t]
\caption{The mean and median (standard error) of spatial ACC of XGBoost and Lasso with and without including the SubX forecasts in their feature sets. Their spatial ACC have improved significantly when the SubX forecasts are included.}
\label{tab: subx_ml}
\centering
\begin{tabular}{|c||c|c||c|c|}
\hline
Features & w/o GMAO  & with GMAO     & w/o NCEP  & with NCEP     \\ \hline
\multicolumn{5}{|c|}{XGBoost}                                                \\ \hline
Mean         & 0.09   (0.02) & {\bf 0.13}   (0.02) & 0.15   (0.02) & {\bf 0.18}   (0.02) \\ \hline
Median       & 0.12   (0.03) &  {\bf 0.14}   (0.04) & 0.21   (0.03) & {\bf  0.23}   (0.02) \\ \hline
\multicolumn{5}{|c|}{Lasso}                                                  \\ \hline
Mean         & 0.12   (0.03) &  {\bf 0.16}   (0.03) & 0.19   (0.01) & {\bf  0.23}   (0.02) \\ \hline
Median       & 0.16   (0.04) & {\bf 0.18}   (0.04) & 0.21   (0.02) & {\bf  0.25}   (0.02) \\ \hline
\end{tabular}
\end{table}

Table \ref{tab: subx_ml} presents the mean and median and their standard error of the spatial ACC using XGBoost and Lasso, with and without the inclusion of SubX forecasts in the feature set. Temporal results are shown in Figure \ref{fig: ml_subx_gmao}. Overall adding either GMAO-GEOS or NCEP-CFSv2 forecasts in the feature set leads to a significant enhancement of predictive skill. We conduct the sign test introduced in \cite{delsole2016forecast} to compare differences in forecast skills. Overall, comparison of ML model performance with and without SubX features yields $p$ values much smaller than 0.01. The one exception is the spatial ACC for Lasso with and without GMAO-GEOS forecasts, for which the $p$ value is 0.21. Furthermore, as shown in Figure \ref{fig: ml_subx_gmao}, the combination of the ML models and the SubX forecasts effectively converts some negative temporal ACC to positive, and strengthens the forecasts originally achieving positive temporal ACC. The improvement is particularly outstanding for the west-north-central region, a region considered hard to predict. Similarly, regarding temporal relative $R^2$, both ML models obtain some improvments in the areas originally characterized by negative values. Especially for Lasso, it picks the central area where the GMAO-GEOS model performs well and obtains positive temporal relative $R^2$. These results highlight the potential to further increase predictive skill of the ML models by incorporating SubX forecasts. We anticipate that more hindcast data from SubX models would lead to notable improvement in predictive skills of the ML models. 

\section{Discussion \& Conclusions}
\label{sec: conc}

In this paper, sub-seasonal climate forecasting, an important but challenging scientific problem, is introduced to the machine learning community. We perform a rigorous evaluation and comparison 
between state-of-the-art machine learning models and two dynamical models from the SubX project, i.e., GMAO-GEOS and NCEP-CFSv2, for SSF in the western contiguous U.S. Experimental results demonstrate that, in general, the ML models 
can outperform the SubX models. However, the ML model forecasts usually are relatively conservative compared to the SubX forecasts which, when correctly made, match the scale of the ground truth better. Acknowledging the strengths of both ML and dynamical models, we obtain significant improvements in predictive skill by including the SubX forecasts as a new feature of ML models, which illustrates the potential in generating skilful SSF by combining such two types of models. Further, we show that ML models make most of bad forecasts during weather extremes, e.g., unusual cold waves, and suggest ways of further improving the ML models by  separately  modeling  extreme  events.

\section*{Acknowledgements}

The research was supported by NSF grants OAC-1934634, IIS-1908104, IIS-1563950, IIS-1447566, IIS-1447574, IIS-1422557, CCF-1451986. The authors would like to acknowledge the high-performance computing support from Casper provided by NCAR's Computational and Information Systems Laboratory, sponsored by the National Science Foundation.

\bibliographystyle{plain}
\bibliography{ref}

\end{document}